\documentclass[12pt]{iopart}

\usepackage{epsfig,graphicx,dcolumn}

\begin{document}

\title{Fitness-driven deactivation in network evolution}

\author{Xin-Jian Xu$^{1,2}$, Xiao-Long Peng$^{1}$, Michael Small$^{3}$ and Xin-Chu Fu$^{1,2}$}

\address{$^{1}$Department of Mathematics, College of Science,
Shanghai University, Shanghai 200444, China\\
$^{2}$Institute of Systems Science, Shanghai University, Shanghai
200444, China\\
$^{3}$Department of Electronic and Information Engineering, Hong
Kong Polytechnic University, Hung Hom, Kowloon, Hong Kong SAR,
China}

\ead{xinjxu@yahoo.com, xlpeng@shu.edu.cn, ensmall@polyu.edu.hk and
xcfu@shu.edu.cn}

\begin{abstract}
Individual nodes in evolving  real-world networks typically
experience growth and decay --- that is, the popularity and
influence of individuals peaks and then fades. In this paper, we
study this phenomenon via an intrinsic nodal fitness function and an
intuitive aging mechanism. Each node of the network is endowed with
a fitness which represents its activity. All the nodes have two
discrete stages: active and inactive. The evolution of the network
combines the addition of new active nodes randomly connected to
existing active ones and the deactivation of old active nodes with
possibility inversely proportional to their fitnesses. We obtain a
structured exponential network when the fitness distribution of the
individuals is homogeneous and a structured scale-free network with
heterogeneous fitness distributions. Furthermore, we recover two
universal scaling laws of the clustering coefficient for both cases,
$C(k) \sim k^{-1}$ and $C \sim n^{-1}$, where $k$ and $n$ refer to
the node degree and the number of active individuals, respectively.
These results offer a new simple description of the growth and aging
of networks where intrinsic features of individual nodes drive their
popularity, and hence degree.
\end{abstract}

\noindent{\it Keywords}: random graphs, networks, fitness, power law

\maketitle

\section{Introduction}

Over the past decade, there has been an explosion of interest in
complex networks for describing structures and dynamics of complex
systems. Despite differences in their nature, many networks may be
characterized by similar topological properties. For instance, real
networks display highly clustering than expected from classic random
graphs \cite{DJW99}. Also, it has been widely observed that
node-degree distributions of many large networks are heavy tailed
\cite{ASBS00}, e.g., exponential and power-law. To understand how
these phenomena arise, research devoted to evolving networks has
rapidly flourished \cite{BA99,KRL00,DMS00,AB00,FM01,KR01}. The basic
premise is that the network will continue to grow at a constant rate
and new nodes attach to old ones with some possibility. When the
newly added nodes connect with equal probability to nodes already
present in the network, the degree distribution of the nodes of the
resulting network is exponential. Whereas for newcomers connecting
to old ones with linear preference of the node degree, Barab\'{a}si
and Albert (BA) observed a power-law distribution of connectivity
\cite{BA99}.

In the real world, agents, represented by nodes, always age after
growth. For instance, in scientific citation networks there is a
half-life effect: old papers are rarely cited since they are no
longer sufficiently topical (or they are more often referenced
through secondary literature). On the World Wide Web popular web
sites (for example, search engines) will often loose favor to newer
alternatives.  To study the effect of this phenomenon on network
evolution, the BA model has been modified by incorporating time
dependence in the network
\cite{DM00,ZWZ03,HS04,HAB07,LR07,CM09,KE02a,KE02b,VA03,WXW05,TL06}.
Dorogovtsev and Mendes studied the case when the connection
probability of the new node with an old one is not only proportional
to the degree $k$ but also to a power of its present age
$\tau^{-\alpha}$ (where $\tau$ is the age of a node) \cite{DM00}.
They found that the network shows scale-free (SF) behavior only in
the region $\alpha < 1$. For $\alpha > 1$, the distribution $P(k)$
is exponential. Yet, the gradual aging model show lower clustering
than realistic networks. On the contrary, Klemm and Egu\'{i}luz
considered evolving networks based on the memory of nodes and
proposed a degree-dependent deactivation network model \cite{KE02a}
which is highly clustered and retains the power-law distribution of
the degree (but no consideration of exponential degree
distributions).

The aim of this paper is to propose a simpler and more fundamental
mechanism to build networks, including SF networks, while retaining
the positive features of aforesaid models. The mechanism we propose
simulates realistic networks, and can be understood analytically. It
has been shown that in some networks, the popularity (or activity)
of a node is essentially determined by the so-called \lq\lq
fitness\rq\rq \cite{BB01,CCRM02,GL04,SC04} which is intrinsically
related to the role played by each node, such as the innovation of a
scientific paper or the content of a webpage. This allows us to
represent the activities of individuals by intrinsic fitnesses and
suggest a fitness-driven deactivation approach to build structured
networks. Compared with topological information, the intrinsic
fitness provides a more natural and appropriate deactivation
criterion for aging of nodes. We show that depending on the
node-fitness distributions two topologically different networks can
emerge, the connectivity distribution following either an
exponential or a generalized power law. In both cases, the networks
are highly clustered. Irrespective of the fitness distribution, we
observe two scaling laws of the clustering coefficient, $C(k) \sim
k^{-1}$ and $C \sim n^{-1}$, where $k$ is the node degree and $n$
corresponds to the number of active individuals in the network.
Hence, this mechanism offers an explanation for the origin and
ubiquity of such clustering in real networks.

\section{Model}

Rather than connectivity-dependent deactivation dynamics of the
nodes developed in \cite{KE02a,KE02b,VA03,WXW05,TL06}, the present
deactivation model is based on the individual fitnesses. For each
node $i$ a fitness $x_i>0$, the random number drawn from a given
probability distribution function $\rho(x)$, is assigned to measure
its popularity or activity. The deactivation mechanism is
characterized by the transition of a node from the active to the
inactive state interpreted as a collective forgetting of it.

The network starts from an initial seed of $n$ nodes, totally
connected by undirect edges, which are all active. Then at each time
step the dynamics runs as follows.

(i) Add a new node $i$, which connects to $m$ ($m \le n$) nodes
randomly chosen from the $n$ active ones. By $k^{{\rm in}}$ we
denote the in-degree of a node, i.e., the number of edges pointing
to it. The in-degree of the newcomer is $k_{i}^{\rm{in}}=0$ at
first. Each selected active node $j$ receives exactly one incoming
edge, thereby $k_{j}^{\rm{in}} \rightarrow k_{j}^{\rm{in}} +1$.
Since the out-degree of each node is $m$ always, the total degree of
a node is $k=k^{\rm{in}}+m$.

(ii) Activate the new node and deactivate one (denoted by $j$) of
the $n$ old active nodes with probability
\begin{equation}
\pi(x_j)=\sigma x_{j}^{-1}, \label{deactprob}
\end{equation}
where $\sigma=(\sum_{j\in{\rm \Lambda}}x_{j}^{-1})^{-1}$ is the
normalization factor. The summation runs over the set $\Lambda$ of
the $n$ old active nodes.

During evolution, a node might receive edges while it is active, and
once inactive it will not receive edges any more. Note that the
fitter the individual is, the more difficult for it to be
deactivated. For the case of the citation network,
Eq.~(\ref{deactprob}) means that the famous paper with great
innovation is less possibility to be forgotten.

\section{Degree distribution}

Denoting by $a(k,x,t)$ the probability of active nodes with degree
$k$ and fitness $x$ at time $t$, we can write out the master
equation for network evolution
\begin{equation}
a(k+1,x,t+1) = a(k+1,x,t)[1-\pi(x)](1-\frac{m}{n}) +
a(k,x,t)[1-\pi(x)]\frac{m}{n}. \label{mastereq}
\end{equation}
At each time step, an active node is deactivated and a newcomer
joins the set $\Lambda$ to keep the number $n$ unchanged. According
to Eq.~(\ref{deactprob}), the normalization factor $\sigma$ varies
with time because of the change of the nodes in the active set.
However, the fitness of each node is a random number taken from a
given probability distribution $\rho(x)$, therefore the
normalization $\sigma$ fluctuates very slightly and can be treated
as a constant. Substituting Eq.~(\ref{deactprob}) into
Eq.~(\ref{mastereq}) yields
\begin{equation}
a(k+1,x,t+1) \approx a(k+1,x,t)(1-\frac{\sigma}{x})(1-\frac{m}{n}) +
a(k,x,t)(1-\frac{\sigma}{x})\frac{m}{n}. \label{mastereqr}
\end{equation}
Imposing the stationarity condition $a(k,x,t)=a(k,x)$, we obtain the
equation
\begin{equation}
a(k+1,x) = \frac{m(x-\sigma)}{m(x-\sigma)+n\sigma}a(k,x) = \left[
\frac{m(x-\sigma)}{m(x-\sigma)+n\sigma} \right]^{k-m}a(m,x)
\end{equation}
for the probability of active nodes with degree $k+1$ and fitness
$x$ in the stationary state, where
\begin{equation}
a(m,x)=\frac{n(x-\sigma)}{m(x-\sigma)+n\sigma}\rho(x)
\end{equation}
is the stationary probability of active nodes with degree $m$ and
fitness $x$. Then we obtain
\begin{equation}
a(k,x) = \frac{n}{m} \left[ \frac{m(x-\sigma)}{m(x-\sigma)+n\sigma}
\right]^{k-m} \rho(x).
\end{equation}
Denoting by $a(k)$ the possibility of active nodes with degree $k$
in the steady state, we have
\begin{equation}
a(k) = \int_0^{x_{{\rm max}}} a(k,x){\rm d}x = \int_0^{x_{{\rm
max}}} \frac{n}{m} \left[ \frac{m(x-\sigma)}{m(x-\sigma)+n\sigma}
\right]^{k-m} \rho(x)\rm{d}x.
\end{equation}
In case the total number $N$ of nodes in the network is larger than
the number $n$ of active nodes, the degree distribution $P(k)$ can
be approximated by considering inactive nodes only. Thus, $P(k)$ can
be calculated as the rate of the change of $a(k)$,
\begin{eqnarray}
P(k) &=& -\frac{{\rm d}a(k)}{{\rm d}k} \nonumber\\
&=& \int_0^{x_{{\rm max}}} \frac{n}{m} \left[
\frac{m(x-\sigma)}{m(x-\sigma)+n\sigma} \right]^{k-m} \ln \left[
\frac{m(x-\sigma)}{m(x-\sigma)+n\sigma} \right]^{-1} \rho(x) {\rm
d}x. \label{integralpk}
\end{eqnarray}
Even when the form of $\rho(x)$ is given, it is still difficult to
solve the integral on the right-hand side of the equation. Instead,
we need a more subtle technique. We assume that
\begin{eqnarray}
F(x) &=& \frac{m(x-\sigma)}{m(x-\sigma)+n\sigma}, \label{deff}\\
G(x) &=& \frac{n}{m} [F(x)]^{-m} \ln[F(x)]^{-1}\rho(x). \label{defg}
\end{eqnarray}
Without lack of generality, we also normalize the fitnesses. Now Eq.
(\ref{integralpk}) can be rewritten as
\begin{equation}
P(k) = \int_0^{1} [F(x)]^{k} G(x) {\rm d}x. \label{integralpks}
\end{equation}
As will be seen below, for the proper choice of $F$ and $G$, one can
construct networks with exponential or power-law degree
distributions, and then determine the forms of the corresponding
fitness distributions.

(i) {\em Exponential degree distribution.} We set $F(x)=1/\mu$ and
$G(x)=\nu$, where $\mu (>1)$ and $\nu$ are positive constants.
Consequently, the integral of Eq. (\ref{integralpks}) is
\begin{equation}
P(k)=\nu\mu^{-k}, \label{pkexp}
\end{equation}
following an exponential. According to the definition of $F(x)$, we
have
\begin{equation}
\frac{m(x-\sigma)}{m(x-\sigma)+n\sigma} = \frac{1}{\mu},
\end{equation}
the solution of which reads
\begin{equation}
x = \sigma + \frac{n\sigma}{m(\mu-1)} = {\rm constant}.
\label{expressx}
\end{equation}
Then the normalization factor becomes
\begin{equation}
\sigma = \left\{ n \left[\sigma +
\frac{n\sigma}{m(\mu-1)}\right]^{-1} \right\}^{-1},
\end{equation}
yielding
\begin{equation}
\mu = 1 + \frac{n}{m(n-1)}.
\end{equation}
According to the definition of $G(x)$, we have
\begin{equation}
\nu = \frac{n}{m} \mu^{m} \ln\mu \rho(x),
\end{equation}
which suggests
\begin{equation}
\rho(x)= \frac{m\nu}{n\mu^{m}\ln\mu} = {\rm constant}.
\label{expressrho}
\end{equation}
According to the normalization, $\rho(x)$ should satisfy
$\int_{0}^{1} \rho(x){\rm d}x =1$. Combining Eqs.~(\ref{expressx})
and (\ref{expressrho}), we finally obtain the distribution of
$\rho(x)$,
\begin{equation}
\rho(x)=\delta\left(x-\frac{\sigma(m\mu-m+n)}{m(\mu-1)}\right).
\end{equation}
That is to say, the fitness of each node is identical. Considering
the constant approximation of $\sigma$, the above result can be
generalized to homogeneous distributions of fitnesses. Here,
homogeneity implies that the node fitnesses have small fluctuations
around the mean $\langle x \rangle$, hence the small variance. We
conclude that for any given $\rho(x)$ distributing homogeneously,
the fitness-driven deactivation generates structured exponential
networks.

(ii) {\em Power-law degree distribution.} We set $F(x)=e^{-\phi x}$
and $G(x)=\varphi x^{\gamma}$, where $\phi$, $\varphi$ and $\gamma$
are positive constants. Accordingly, Eq. (\ref{integralpks}) can be
rewritten as
\begin{equation}
P(k)=\frac{\varphi}{(\phi k)^{\gamma+1}} \int_{0}^{\phi k}
e^{-z}z^{\gamma}{\rm d}z. \label{integralpkpow}
\end{equation}
In the limit $k \rightarrow \infty$, one has $\int_{0}^{\phi k}
e^{-z}z^{\gamma}{\rm d}z=\Gamma(\gamma+1)$. For any finite $k$, the
integral $\int_{0}^{\phi k} e^{-z}z^{\gamma}{\rm d}z$ is convergent
and smaller than $\Gamma(\gamma+1)$. Thus the connectivity
distribution has a power-law form
\begin{equation}
P(k) \sim k^{-\gamma-1}. \label{pkpow}
\end{equation}
Combining Eqs. (\ref{deff}) and (\ref{defg}), we obtain
\begin{equation}
\rho(x) = \frac{\varphi m}{\phi n}x^{\gamma-1}e^{-\phi mx}.
\end{equation}
This heavy tailed distribution implies large fluctuations of the
node fitnesses, hence heterogeneity. We conclude that the
fitness-driven deactivation can build structured SF networks if the
node fitnesses are distributed heterogeneously.

\begin{figure}
\includegraphics[width=\columnwidth]{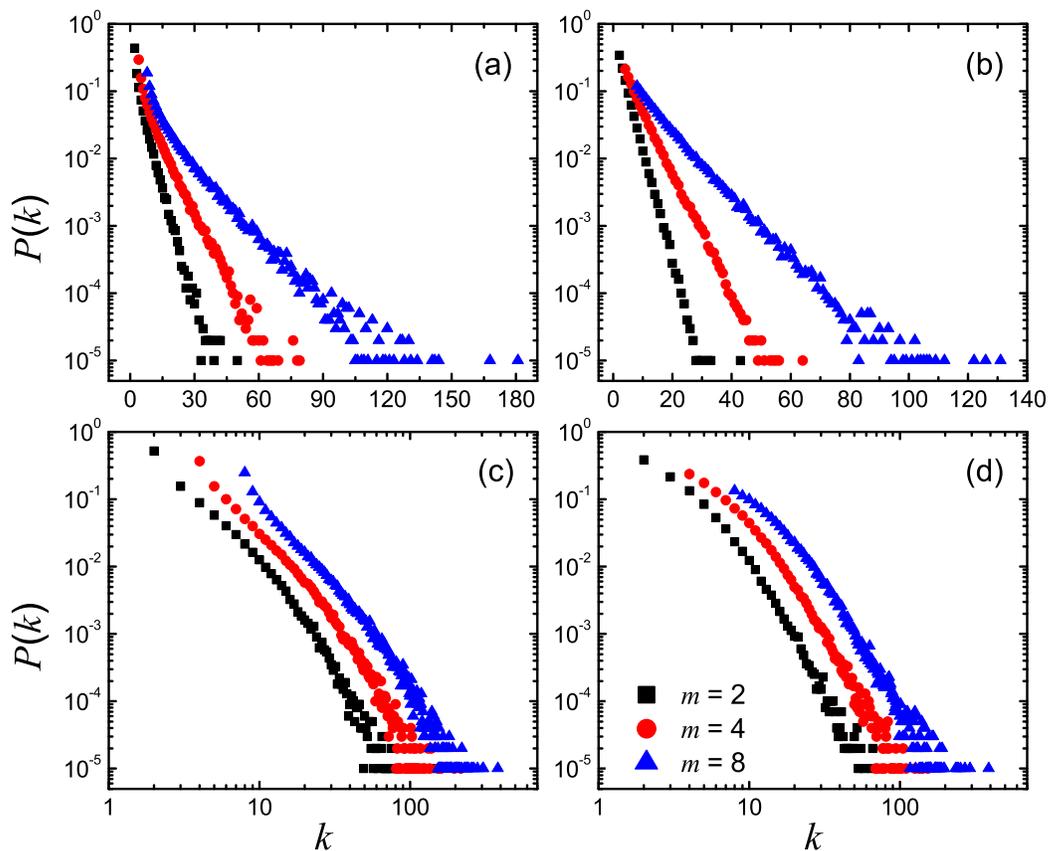}
\caption{(Color online) Degree distributions of nodes of generated
networks for different fitness distribution functions: uniform (a),
Gaussian (b), exponential $e^{-x}$ (c), and power-law $x^{-1}$ (d).
All the fitnesses have been normalized. For uniform and Gaussian
distributions with the same mean ($0.5$) and small variances
($0.08(4)$ and $0.01(6)$), the node fitnesses fluctuate slightly
around the mean and can be regarded as the homogeneity; whereas for
exponential and power-law distributions, there are large
fluctuations of the node fitnesses due to the right-skewed feature,
resulting in the heterogeneity. The experiment networks start from
the initial $n=10$ nodes and end with the total population
$N=10^5$.} \label{fig1}
\end{figure}

\begin{figure}
\includegraphics[width=\columnwidth]{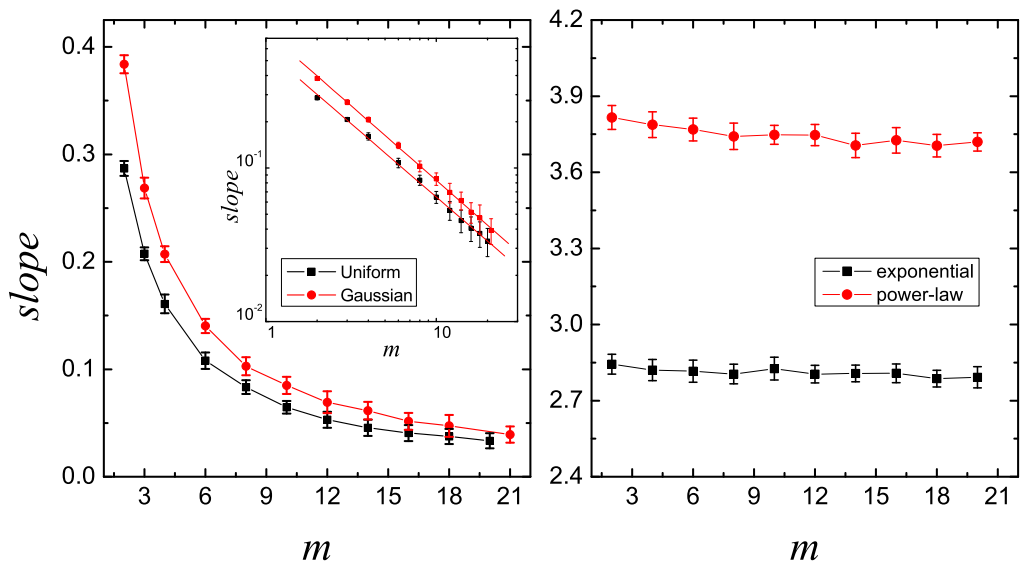}
\caption{(Color online) The dependence of the slopes of the lines in
Fig.~\ref{fig1} on $m$. In the inset, the data reported on the
log-log representation show a power law $m^{-\xi}$, and the best
linear fit gives $\xi=0.98(1)$.} \label{fig2}
\end{figure}

In Fig.~\ref{fig1} we give the simulation results of degree
distributions $P(k)$ for four kinds of distribution functions
$\rho(x)$ of node fitnesses. In case that the distributions of
fitnesses are homogeneous, e.g., uniform (Fig. \ref{fig1}(a)) and
Gaussian (Fig. \ref{fig1}(b)), the linear-log plots imply an
exponential degree distribution. Conversely, when the fitnesses
distribute heterogeneously, the log-log plots of Figs. \ref{fig1}(c)
and \ref{fig1}(d), corresponding to exponential and power-law cases
respectively, predict a generalized power law. We obtain the slopes
of the lines in Fig.~\ref{fig1} by least squares fitting and plot
them as a function of $m$ in Fig.~\ref{fig2}. For homogeneous
fitnesses, we notice a scaling relation between the slope and $m$,
whereas for heterogenous fitnesses, the slope shows independence on
$m$.

\section{Clustering coefficient}

In a network, if a node $i$ has $k_i$ edges, and among its $k_i$
nearest neighbors there are $e_i$ edges, then the clustering
coefficient of $i$ is defined by
\begin{equation}
c_i=\frac{2e_i}{k_i(k_i-1)}. \label{defclusteri}
\end{equation}
In the deactivation model, new edges are created between the
selected active nodes and the added one. Let us first consider the
case of $n=m$. At each time step, the degree $k_i$ of the active
node $i$ increases by $1$ and $e_i$ increase by $m-1$ until it is
deactivated. Therefore, the evolutionary dynamics of $k_i$ and $e_i$
are given by
\begin{eqnarray}
k_i &=& (m+t), \label{evolutionk}\\
\frac{de_i}{dt} &=& (m-1). \label{evolutione}
\end{eqnarray}
Integrating Eq. (\ref{evolutione}) with the boundary condition
$e_i(0)=m(m-1)/2$ and substituting the solution into Eq.
(\ref{defclusteri}), we recover the clustering coefficient $c(k)$
restricted to the nodes of degree $k$ \cite{KE02b,VA03}
\begin{equation}
c(k)=\frac{2(m-1)}{k-1}-\frac{m(m-1)}{k(k-1)}. \label{expressck1}
\end{equation}
For $n>m$, the clustering coefficient $C(k)$ is just the
generalization of Eq. (\ref{expressck1}),
\begin{equation}
C(k)=\frac{m}{n} \left[ \frac{2(m-1)}{k-1}-\frac{m(m-1)}{k(k-1)}
\right], \label{expressck2}
\end{equation}
which indicates that the local clustering scales as $C(k) \sim
k^{-1}$ for large $k$. The clustering coefficient $C$ of the whole
network is the mean of $C(k)$ with respect to the degree
distribution $P(k)$,
\begin{equation}
C = \int_m^{\infty} C(k)P(k) {\rm d}k. \label{defcluster}
\end{equation}
Substituting Eqs. (\ref{pkexp}) and (\ref{pkpow}) into the integral
respectively, we obtain
\begin{equation}
C \sim \left\{
\begin{array}{lll}
& \frac{2\nu(m^2-m)}{n}\int_m^{\infty}\frac{\mu^{-k}}{k}{\rm d}k+\frac{\nu(m^2-m^3)}{n}\int_m^{\infty}\frac{\mu^{-k}}{k(k-1)}{\rm d}k,\\
& \frac{2(m^2-m)}{n}\int_m^{\infty}\frac{{\rm d}k}{k^{\gamma+2}}+\frac{(m^2-m^3)}{n}\int_m^{\infty}\frac{{\rm d}k}{k^{\gamma+2}(k-1)}, \\
\end{array}
\right. \label{cluster}
\end{equation}
proportional to $n^{-1}$.

\begin{figure}
\includegraphics[width=\columnwidth]{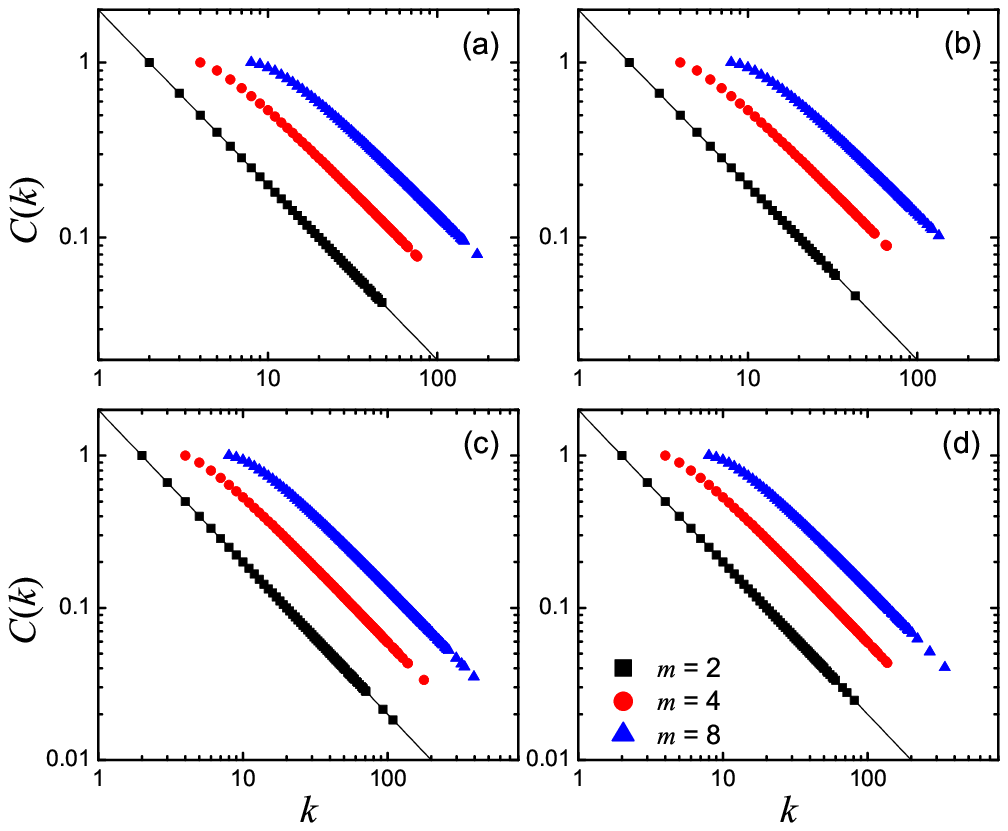}
\caption{(Color online) Log-log plots of $C(k)$ as a function of $k$
for different fitness distribution functions: uniform (a), Gaussian
(b), exponential $e^{-x}$ (c), and power-law $x^{-1}$ (d). The solid
lines correspond to a power law $2k^{-1}$.} \label{fig3}
\end{figure}

\begin{figure}
\includegraphics[width=\columnwidth]{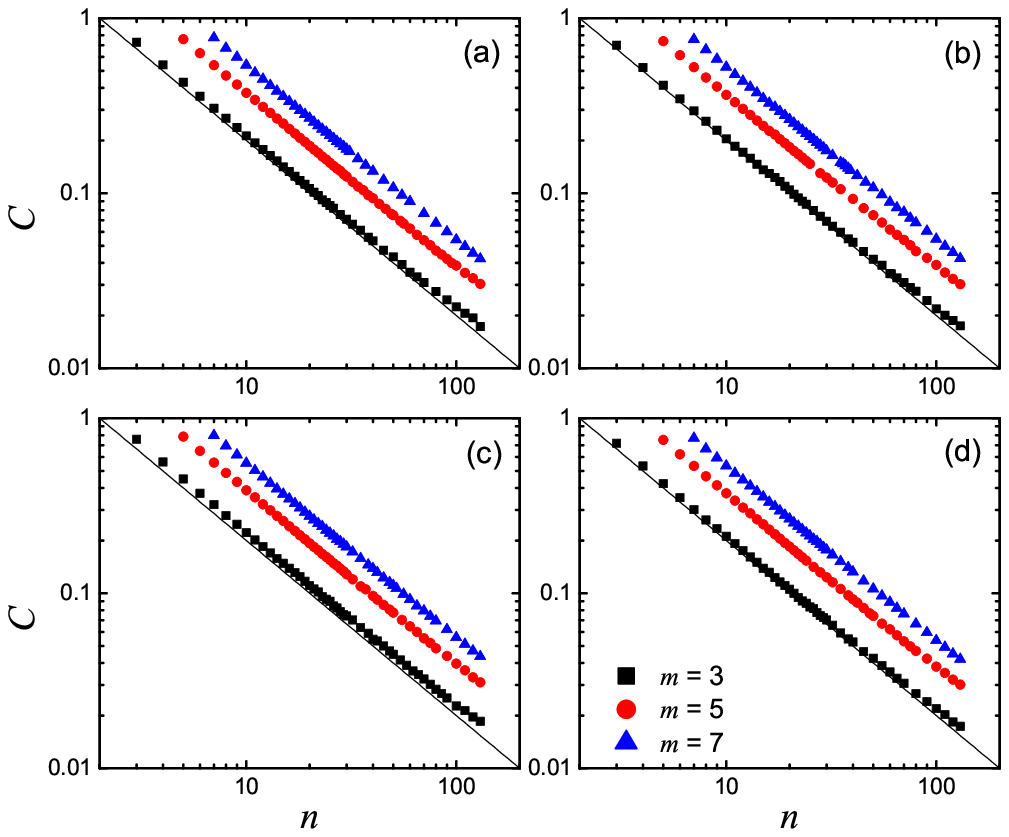}
\caption{(Color online) Log-log plots of $C$ as a function of $n$
for different fitness distribution functions: uniform (a), Gaussian
(b), exponential $e^{-x}$ (c), and power-law $x^{-1}$ (d). The solid
lines correspond to a power law $2n^{-1}$.} \label{fig4}
\end{figure}

In Fig. \ref{fig3} we show the simulation results of $C(k)$ as a
function of $k$. All the plots follow a power law $C(k) \sim
k^{-1}$, which coincides with the expression in Eq.
(\ref{expressck2}). For the clustering coefficient of the whole
network, as shown in Fig. \ref{fig4}, the linearity of all the plots
also implies a power-law relation $C \sim n^{-1}$. It is worth
noting that both the scaling laws are independent of fitness
distribution functions. One obtains the same result for uniform,
Gaussian, exponential and power-law distributions of fitnesses.

\section{Conclusion}

In this paper, we have presented an alternatively simple and
intuitive model for a large and important class of networks widely
observed in the real world. We defined the intrinsic fitness as the
way of quantifying the popularity of individuals, i.e., the fitter a
node is, the higher possibility it is active. The growth dynamics of
the network is governed by the naive fitness-driven deactivation
mechanism. The deactivation probability of a node is proportional to
the inverse of its fitness, which characterizes the individual
capability of obtaining further links. We studied the connectivity
distribution and the clustering coefficient that can fundamentally
shape a network. On one hand, we found the great influence of the
node fitnesses on the connectivity distribution. The homogeneous
fitnesses generate exponential networks, while the heterogeneous
fitnesses result in SF ones. On the other hand, we recovered two
universal scaling laws of the clustering coefficient regardless of
the fitness distributions. These results are consistent with what
has been empirically observed in many real-world networks in (for
example) \cite{WM00,MN01,RB03,BMG04}, and so the present model
provides a new way to understand complex networks with age.

\section*{Acknowledgments}

This work was jointly supported by NSFC (Nos. 10805033 and
11072136), HK UGC GRF (No. PolyU5300/09E), and by Shanghai Leading
Academic Discipline Project S30104.

\end{document}